\documentclass{PoS}
\newcommand{\ba}{\begin{eqnarray}}
\newcommand{\ea}{\end{eqnarray}}

\newcommand{\nn}{\nonumber}

\newcommand{\bs}{\bold}
\newcommand{\es}{&=&}

\title{Wigner Distributions Using Light-Front Wave Functions}

\ShortTitle{Wigner Distributions}

\author{\speaker{Asmita Mukherjee}\\
        Department of Physics, Indian Institute of Technology Bombay;
        Powai, Mumbai 400076, India\\ 
        E-mail: \email{asmita@phy.iitb.ac.in}}

\abstract{We report on some recent advances in calculating the Wigner
distributions for quarks and gluons, using overlaps of light-front wave
functions.}

\FullConference{QCD Evolution 2017\\
		22-26 May, 2017\\
		Jefferson Lab Newport News, VA - USA}

\begin{document}

\section{Introduction}

The most challenging task in hadron physics today is to understand the
tomographic picture of the nucleon in terms of quarks and gluons. This leads
to the study of the spatial and momentum distributions of these quarks and
gluons as well as their spin and angular momentum correlations. In this
context, Wigner distributions \cite{Wigner32} are objects of intense interest in recent
days. Six dimensional Wigner distributions of quarks and gluons in the rest
frame of the nucleon were  introduced in \cite{Ji03}. In quantum mechanics,
Heisenberg's uncertainty principle prevents us from having a join position
and momentum space description of a system. So Wigner functions are not
positive definite, or they do not have probabilistic  interpretation. A five
dimensional object in the infinite momentum frame or equivalently in
light-front formalism was introduced in \cite{Lorce11}, which is a function of three
momenta and two transverse position coordinates. Taking a Fourier transform
with respect to the transverse position $b_\perp$ relates the Wigner
distributions to the generalized transverse momentum dependent pdfs (GTMDs)
\cite{Meissner09},
which are the so-called mother distributions of the generalized parton
distributions (GPDs) and transverse momentum dependent pdfs (TMDs). Some of
the Wigner distributions do not have a forward limit related to a known TMD.
In fact these are related to the elusive OAM and spin-orbit correlations of
quarks  and gluons \cite{Lorce11,hatta12}. Although there are quite a few theoretical studies to
probe the GTMDs in experiments \cite{Hatta17,Hatta16,Hagi,Bhattacharya}, there are so far no experimental data on
them. Model calculations of Wigner functions are thus important to get a
qualitative idea of them. In this talk, we report on a few recent
developments on model calculations of the quark and gluon Wigner distributions
\cite{moreq17,moreg17,dipankar1}.

\section{Wigner Distributions in a dressed quark model}\label{WD}

We use light-front Hamiltonian formulation \cite{Hari99} in light-cone gauge, using
overlaps of LFWFs. The Wigner distributions of quarks are defined as follows:
\cite{Meissner09, Lorce11}
\ba
\rho^{[\Gamma]} ({ b}_{\perp},{k}_{\perp},x,s,s') = \int \frac{d^2
\Delta_{\perp}}{(2\pi)^2} e^{-i {\Delta}_{\perp}.{ b}_{\perp}}
W_{s\, s'}^{[\Gamma]} ({ \Delta}_{\perp},{k}_{\perp},x)
\ea
where ${b_\perp}$ is the impact parameter   conjugate to ${
\Delta}_\perp$, which is the  transverse momentum transfer.
The quark-quark correlator $W^{[\Gamma]}$ in the above expression can be
written as, 
\ba
W_{s\,s'}^{[\Gamma]} ({\Delta}_{\perp},{k}_{\perp},x)
&=&\int \!\!\frac{dz^{-}d^{2} {z}_{\perp}}{2(2\pi)^3}e^{i k.z}
\Big{\langle}p^{+},\frac{{\Delta}_{\perp}}{2},s' \Big{|}
\overline{\psi}(-\frac{z}{2})\Omega \Gamma \psi(\frac{z}{2}) \Big{|}
p^{+},-\frac{{\Delta}_{\perp}}{2},s\Big{\rangle }  \Big{|}_{z^{+}=0}
\ea

$P={1 \over 2} (p'+p)$ is the average four-momentum of the target state,
the momentum transfer in the transverse direction is $\Delta=p'-p$. $s$($s'$)
is the helicity of the initial
(final) target state. The average four momentum of the active quark is $k$,
and $x$ is the longitudinal momentum fraction of the
parton. $\Gamma$ is the Dirac matrix and $\Omega$ is the gauge link for color gauge invariance.
We use light-cone gauge and take the
gauge link to be  unity.

The Wigner distribution of the gluon are defined as
\cite{Meissner09, Lorce11}
\ba
x W_{\sigma, \sigma'}(x,{k}_{\perp},{b}_{\perp})\es\int \frac{d^2
{\Delta}_{\perp}}
{(2\pi)^2}e^{-i{\Delta}_{\perp}.{b}_{\perp}} \int
\frac{dz^{-}d^{2} z_{\perp}}{2(2\pi)^3 p^+}e^{i k.z} \nn \\
 &\times&\Big{\langle } p^{+},-\frac{{\Delta}_{\perp}}{2},\sigma' \Big{|}
\Gamma^{ij} F^{+i}\Big( -\frac{z}{2}\Big) F^{+j}\Big( \frac{z}{2}\Big)
\Big{|}
p^{+},\frac{{\Delta}_{\perp}}{2},\sigma  \Big{\rangle }  \Big{|}_{z^{+}=0}
\label{eq1}
\ea
Suppressing the color indices, we have, 
\ba
F^{+i} = \partial^+ A^i -  \partial^i  A^+ + g f^{abc} A^+ A^i
\ea

The operator structure for gluon at twist two are \cite{Lorce13}
(i) $\Gamma^{ij}=\delta_\perp^{ij}$,~~
(ii) $\Gamma^{ij} =-i\epsilon^{ij}_{\perp}$ ~~~
(iii) $\Gamma^{ij}=\Gamma^{RR}$ and 
(iv) $\Gamma^{ij}=\Gamma^{LL}$, where $L(R)$ are left(right) polarization of
the gluon. Wigner distribution
for gluons need two gauge links for color gauge invariance, which we have
not shown explicitly in the above expression. 
We use light-cone gauge and take the gauge links to be unity. $\sigma$ and
$\sigma'$ are the helicities of the target state.  Instead
of a proton target, we take the state to be a quark dressed with a gluon at
one loop in perturbation theory.  This can the thought of as a field theory
inspired perturbative model having a gluonic degree of freedom.  The quark
has non-zero mass and both quark and gluon have non-zero transverse momenta. 
A dressed quark state can be expanded in Fock space as
\ba
  \Big{| }p^{+}, {p}_{\perp}, s \Big{\rangle} &=& \Phi^{s}(p)
b^{\dagger}_{s}(p) | 0 \rangle +
 \sum_{s_1 s_2} \int \frac{dp_1^{+}d^{2}p_1^{\perp}}{ \sqrt{16 \pi^3
p_1^{+}}}
 \int \frac{dp_2^{+}d^{2}p_2^{\perp}}{ \sqrt{16 \pi^3 p_2^{+}}} \sqrt{16
\pi^3 p^{+}}
 \delta^3(p-p_1-p_2) \nn \\[1.5ex]
 &&\times\Phi^{s}_{s_1 s_2}(p;p_1,p_2) b^{\dagger}_{s_1}(p_1)
 a^{\dagger}_{s_2}(p_2)  | 0 \rangle
 \ea
The two-particle light-front wave function (LFWF) is written in terms of
 the boost-invariant LFWF as
\ba\sqrt{P^+}\Phi(p; p_1, p_2) = \Psi(x_{i},{q}_{i}^{\perp})\ea
This can be calculated in light-front Hamiltonian
perturbation theory and is given by:

\ba
\Psi^{sa}_{s_1 s_2}(x,{q}^{\perp})&=& 
\frac{1}{\Big[m^2 - \frac{m^2 + ({q}^{\perp})^2 }{x} - \frac{({
q}^{\perp})^2}{1-x} \Big]}
\frac{g}{\sqrt{2(2\pi)^3}} T^a \chi^{\dagger}_{s_1}
\frac{1}{\sqrt{1-x}}\nn\\
&\times&
 \Big[ -\frac{2{q}^{\perp}}{1-x}   -  \frac{({\sigma}^{\perp}.{
q}^{\perp}){\sigma}^{\perp}}{x}
+\frac{im~{\sigma}^{\perp}(1-x)}{x}\Big]
\chi_s ({\epsilon}^{\perp}_{s_2})^{*}
\ea
We here use  two-component formalism \cite{Zhang93}, $\chi$, $T^a$, $m$ and ${
\epsilon}^{\perp}_{s2}$ are the two-component spinor, color 
SU(3) matrices, mass of the quark, and polarization 
vector of the gluons, respectively. 

The Wigner distributions for different combinations of polarizations of the 
target and the probed quark or gluon are expressed in terms of overlaps of the 
light-front wave functions \cite{moreq17}. The single particle sector of the Fock space 
plays an important
role when $x=1$ through the wave function renormalization.
Wigner distributions are calculated analytically using LFWFs and the Fourier
transform with respect to $\Delta_\perp$ is done numerically. 
The analytic results for all the quark and gluon Wigner distributions for a
simple composite spin $1/2$ system can
be found in \cite{moreq17,moreg17}. The full calculation of the Wigner distributions and
spin-spin correlations in quark-diquark model, where the LFWF is modeled by
Ads/QCD correspondence, can be found in \cite{dipankar1}. 
 Here we report on some of our numerical results.

\section{Numerical Results}
The Wigner distributions are functions of 
five variables, two transverse position ${b}_\perp$, two
transverse momenta 
${k}_\perp$ and one longitudinal momentum fraction for the quark(gluon) $x$. 
In the plots we integrate them over $x$.  
For our numerical calculations, we use a better integration strategy
called the Levin method \cite{Levin82}, which suits our
oscillatory integrands, improving the convergence of the results with
respect to the change in the upper limit of the $\Delta_\perp$ integration
as was seen in the earlier calculation \cite{Asmita14,Asmita15}. 
We have used the upper
limit of the $\Delta_\perp$ integration to be $\Delta_{max} =
20~\mathrm{GeV}$, and the results are independent of this cutoff.
We have taken  the mass of the quark to be $m=0.33 ~GeV$.  
For all plots in $b_\perp$ space, we took a fixed value of $k_\perp$ and
vice versa. 

Fig. 1(a) shows the plot of $\rho_{UU}$ in $b$ space, which is 
the distribution of an unpolarized quark in an unpolarized target state.
Fig. 1(b) shows the plot of $\rho_{UL}$ in $b$ space. For a dressed quark,
 $\rho_{UL}=\rho_{LU}$.
This Wigner distribution is related to the orbital angular momentum of the
quark. A dipole structure similar to chiral quark soliton model and
constituent quark model \cite{Lorce11} is seen. In Fig. 2(a) we have plotted $\rho^x_{UT}$,
which gives the distribution of transversely polarized quark in an unpolarized target;
quark polarization in the $x$ direction. Here we also observe a dipole
structure. TMD limit of $\rho_{UT}$ is the Boer-Mulders function. As we have
taken the gauge link to be unity, we cannot access the T-odd
distributions like the Boer-Mulders function.
The behavior in $b$ space is similar to spectator model \cite{Liu15}.
In Fig. 2(b) we have shown  $\rho^x_{LT}$ in
${k}_\perp$ space. These distributions
describe a transversely polarized quark in a longitudinally
polarized target state and here the direction of the polarization of the
quark is in the $x$-direction. We observe a dipole structure in
$k$-space. The TMD limit of this is related to the worm-gear function,
$h^\perp_{1L}$.  

\begin{figure}[h]
 \centering
(a)\includegraphics[width=6cm,height=4cm]{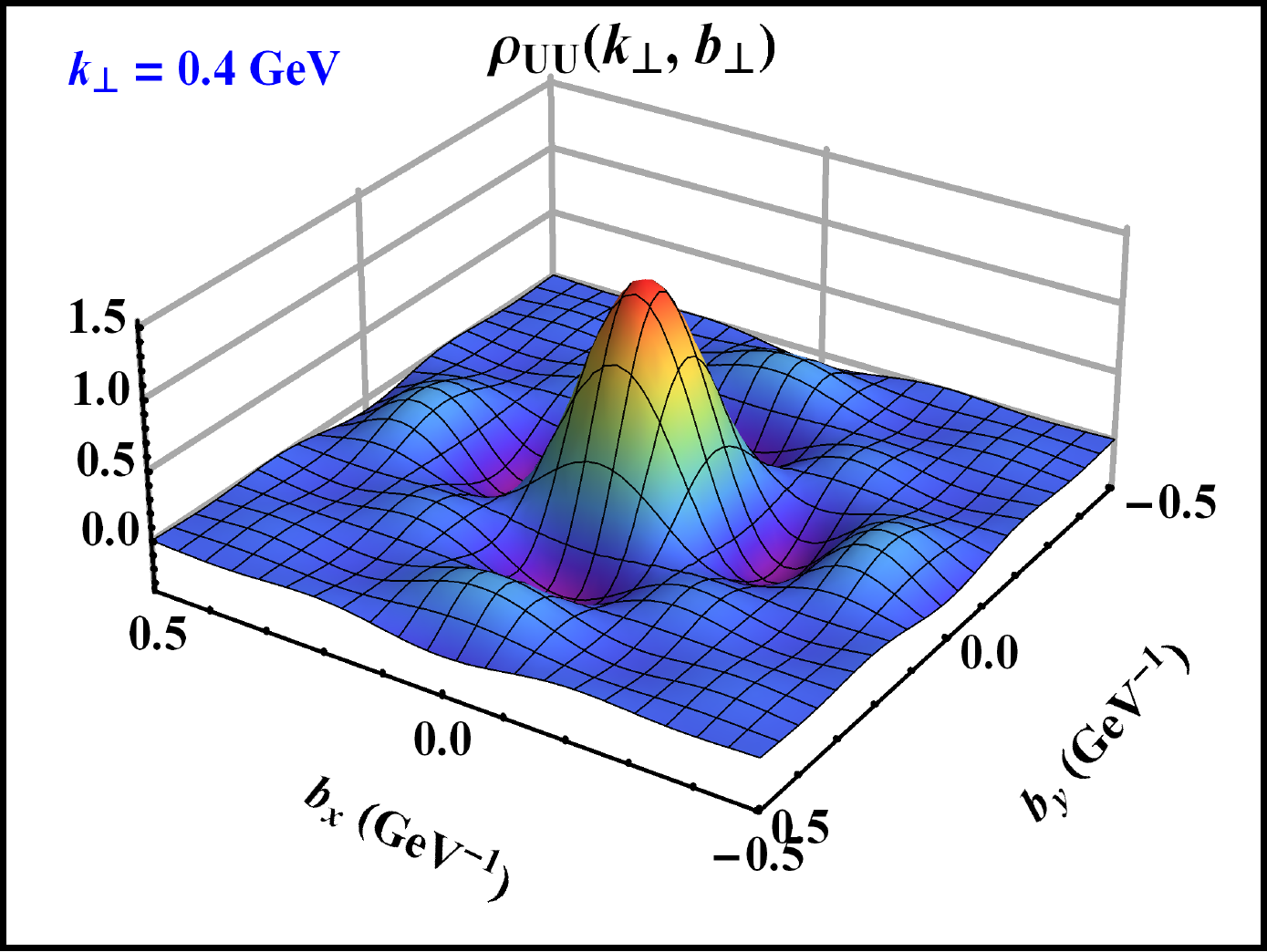}\hskip 0.3cm(b)
\includegraphics[width=6cm,height=4cm]{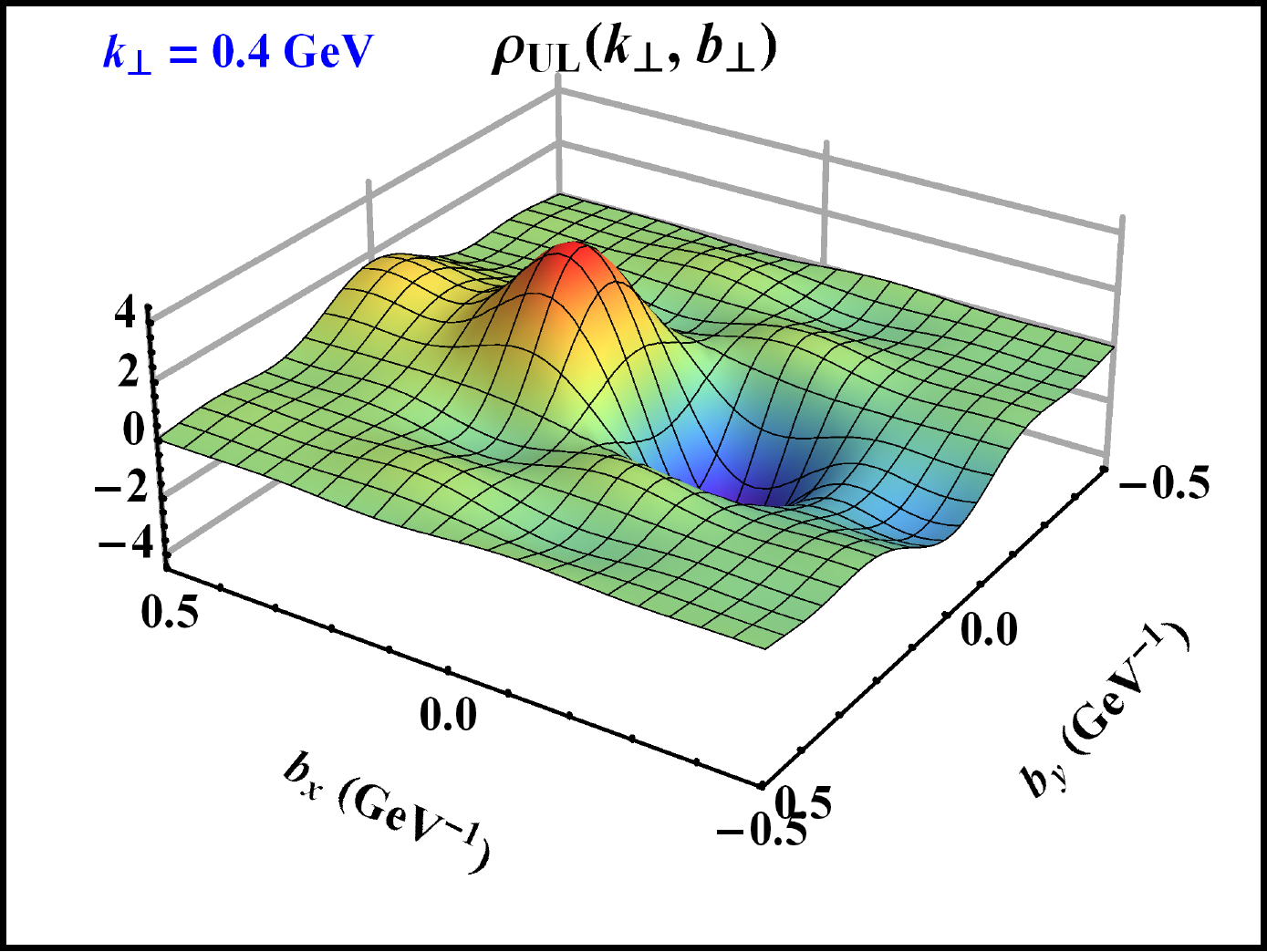}\\[5ex]
 \caption{3D plots of  Wigner distributions $\rho_{UU}({\bs k}_\perp, {\bs
b}_\perp)$ and
  $\rho_{UL}({\bs k}_\perp, {\bs b}_\perp)$ at $\Delta_{max} =
20~\mathrm{GeV}$ in $b$ space \cite{moreq17}.}
\end{figure}

\begin{figure}[h]
 \centering
(a)\includegraphics[width=6cm,height=4cm]{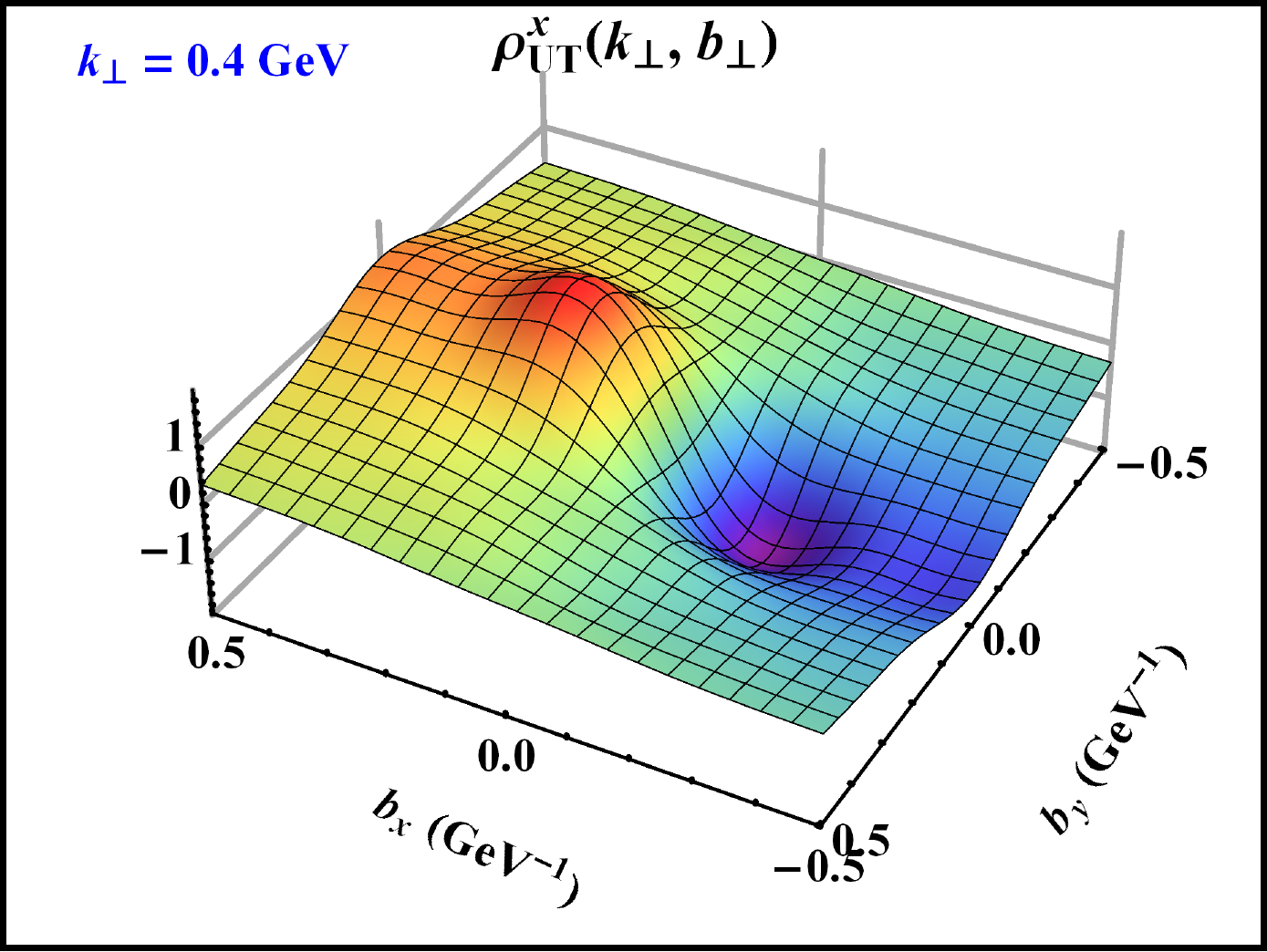}\hskip 0.3cm(b)
\includegraphics[width=6cm,height=4cm]{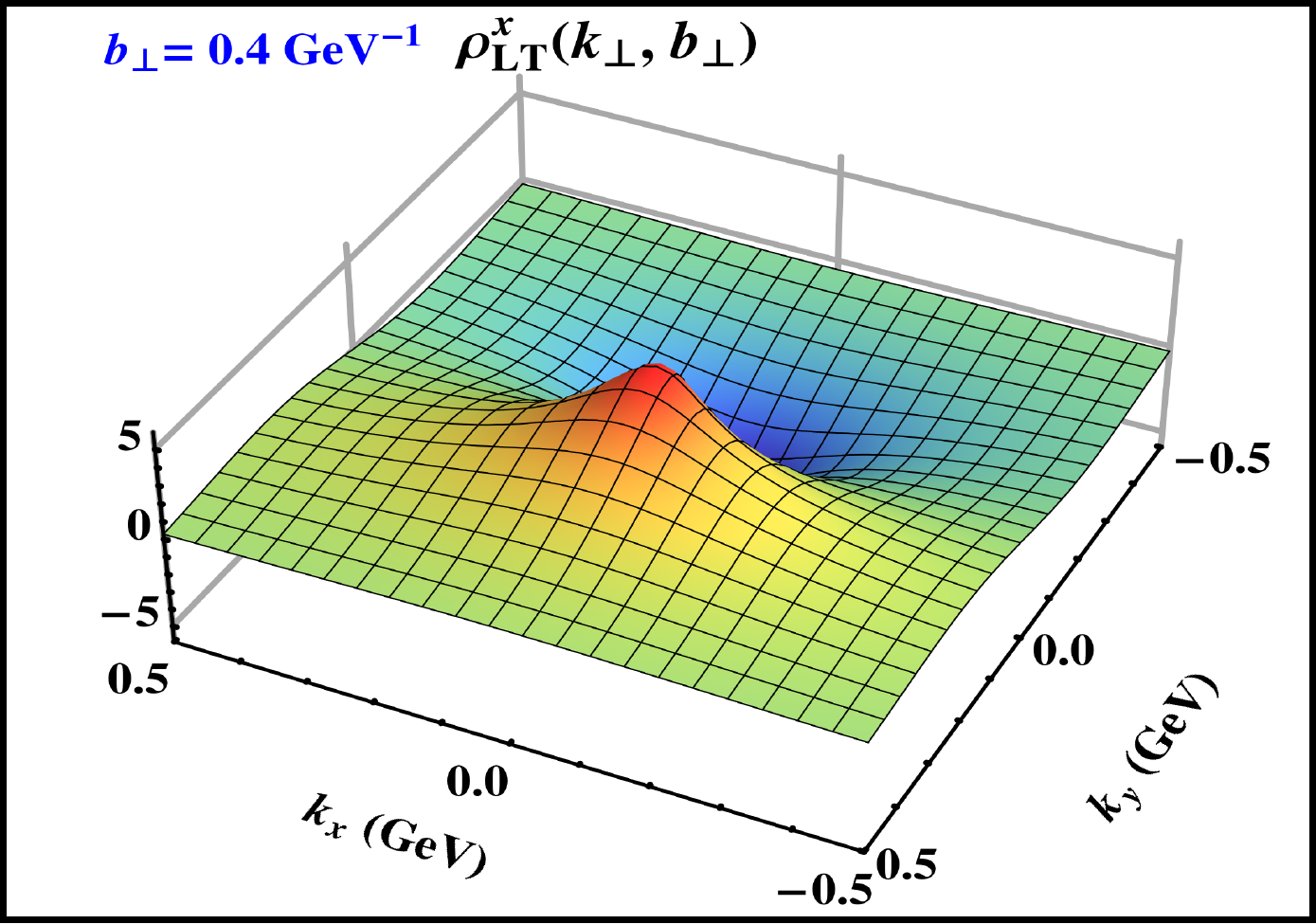}\\[5ex]
 \caption{3D plots of  Wigner distributions $\rho^x_{UT}({\bs k}_\perp, {\bs
b}_\perp)$ in $b$ space and $\rho^x_{LT}({\bs k}_\perp, {\bs
b}_\perp)$ in $k$ space at $\Delta_{max} = 20~\mathrm{GeV}$ \cite{moreq17}.}
\end{figure}

In Fig. 3(a) we have plotted the gluon Wigner distribution $W_{UU}$, which is
the distribution of an unpolarized gluon in an unpolarized dressed quark.
As discussed above, each gluon Wigner distribution needs two
gauge links for color gauge invariance. Depending on whether it is a $++$ or $+-$
gauge link combination, it is called a Weizsacker-Williams type or dipole
type gluon distribution. In \cite{Hatta16} it was shown that both of them
give the same orbital angular momentum distribution of the gluon. In our
calculation, we have taken the gauge links to be unity in light-cone gauge. The
gluon distribution has a positive peak at the center of the $b$ space.
Fig. 3 (b) shows $W_{UL}$,  which is the  Wigner distribution 
for a longitudinally polarized gluon in an unpolarized target state in ${b_\perp}$ 
space. This  shows dipolelike structure. 
Note that the behavior of $W_{UU}$ near $b_\perp=0$ is determined by the
relative dominance  of the $k_\perp^2$ and $\Delta_\perp^2 (1-x)^2$ terms in
the numerator. As $\Delta_{max}$ increases, the second term dominates over
the first, as a result, the peak at $b_\perp=0$ becomes positive.     

\begin{figure}[h]
 \centering
(a)\includegraphics[width=6cm,height=4cm]{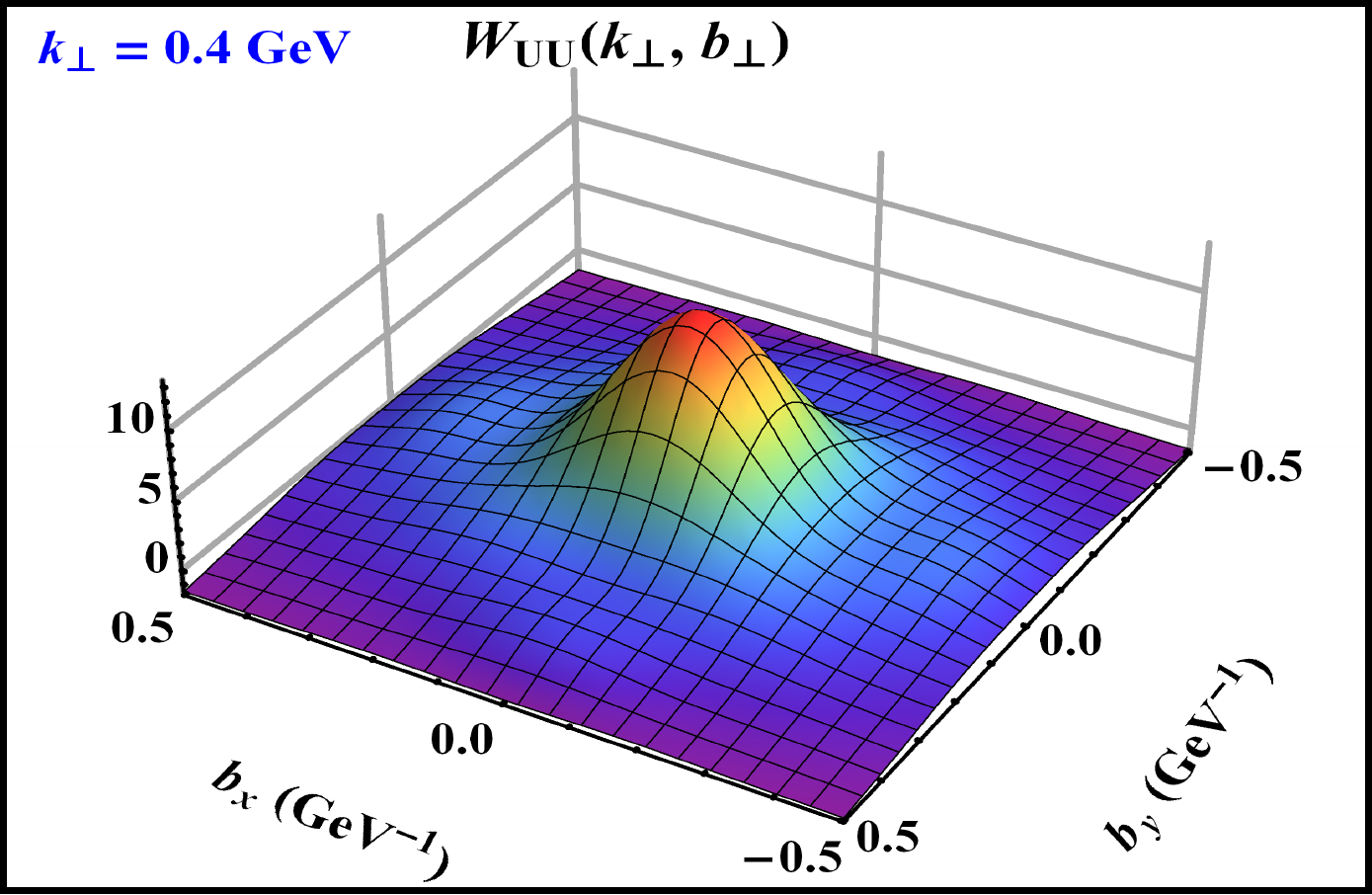}\hskip 0.3cm(b)
\includegraphics[width=6cm,height=4cm]{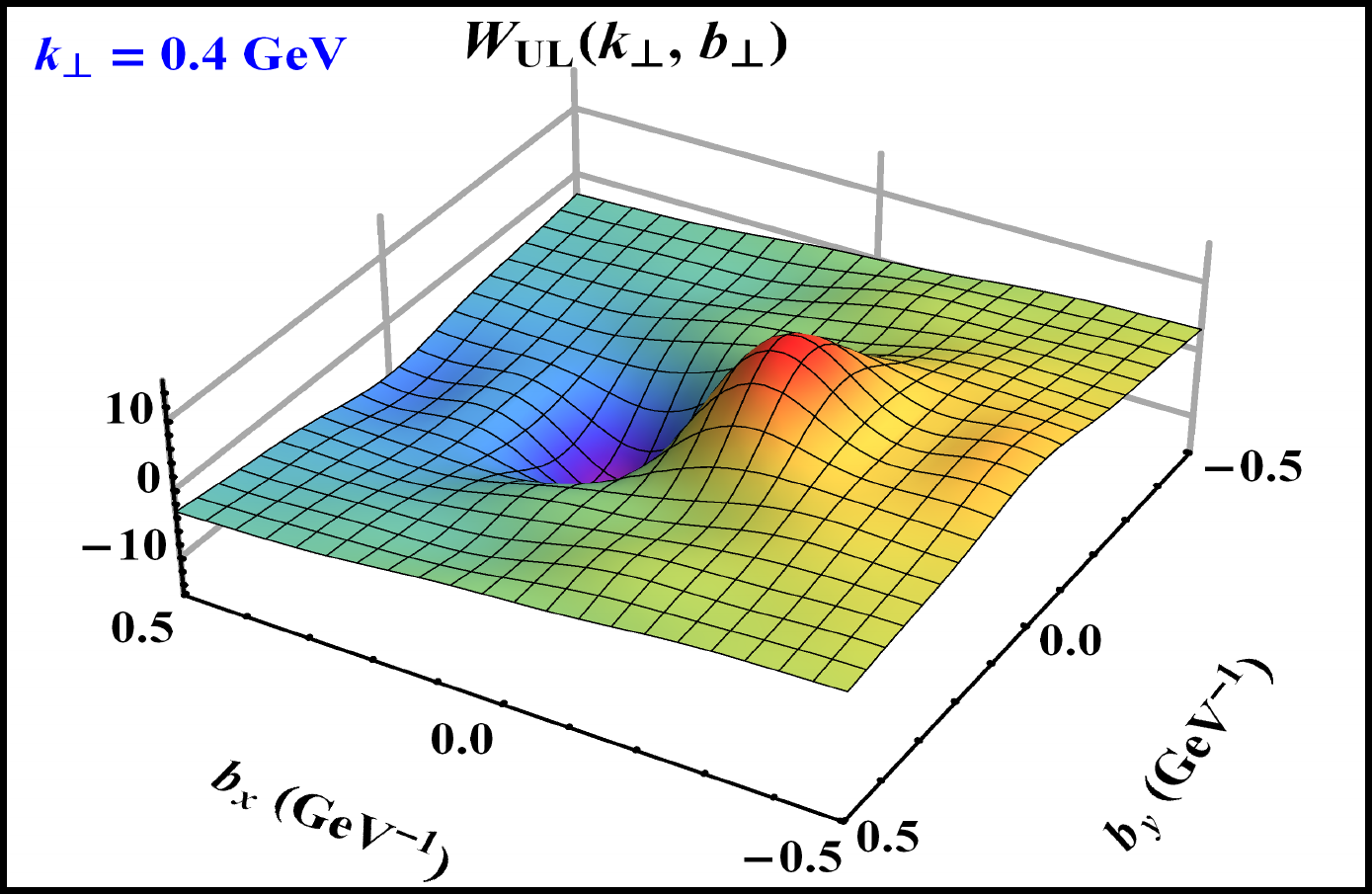}\\[5ex]
 \caption{3D plots of  gluon Wigner distributions $W_{UU}$ and $W_{UL}$ in $b$ space
 for  $\Delta_{max} = 20~\mathrm{GeV}$ \cite{moreg17}.}
\end{figure}

In Fig. 4(a) and 4(b) we have shown the Wigner distributions $\rho_{UU}$ and
$\rho_{UL}$ respectively for the u quark in a quark-diquark model for the
proton, modeled using a modified form of LFWF obtained using Ads/QCD
prediction \cite{dipankar1}.  Here also, we have integrated over $x$. Average quadrupole
distortion for $\rho_{UU}$ is found to be zero in this model. Unlike other
models, no favored direction is there in $b$ space, and the distribution is
circularly symmetric. $\rho_{UL}$ has a dipole structure. The OAM for $u$
quarks is antiparallel to quark spin, same as in scalar diquark model and
opposite to that observed in constituent quark model.   

\begin{figure}[h]
 \centering
(a)\includegraphics[width=6cm,height=4cm]{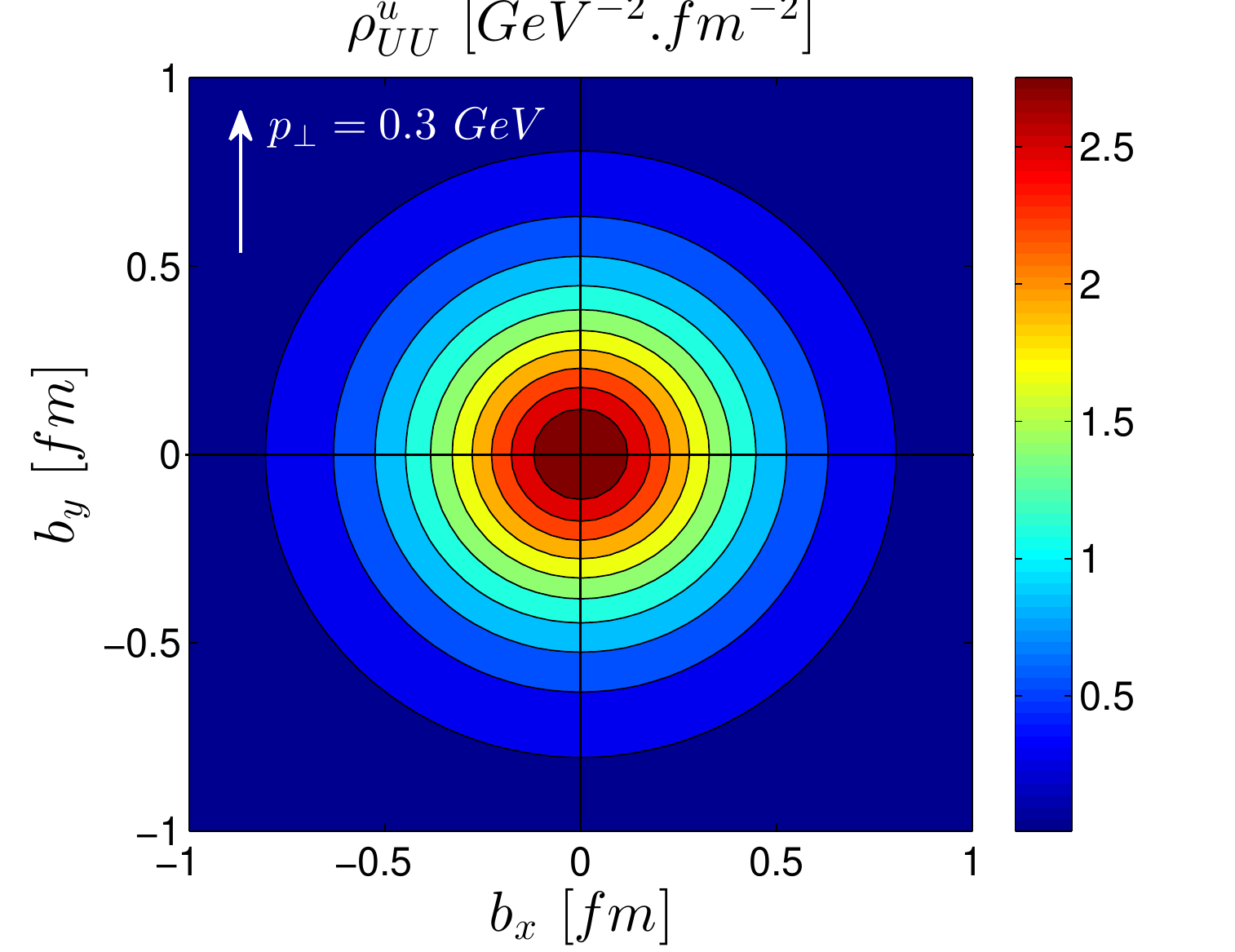}\hskip 0.3cm(b)
\includegraphics[width=6cm,height=4cm]{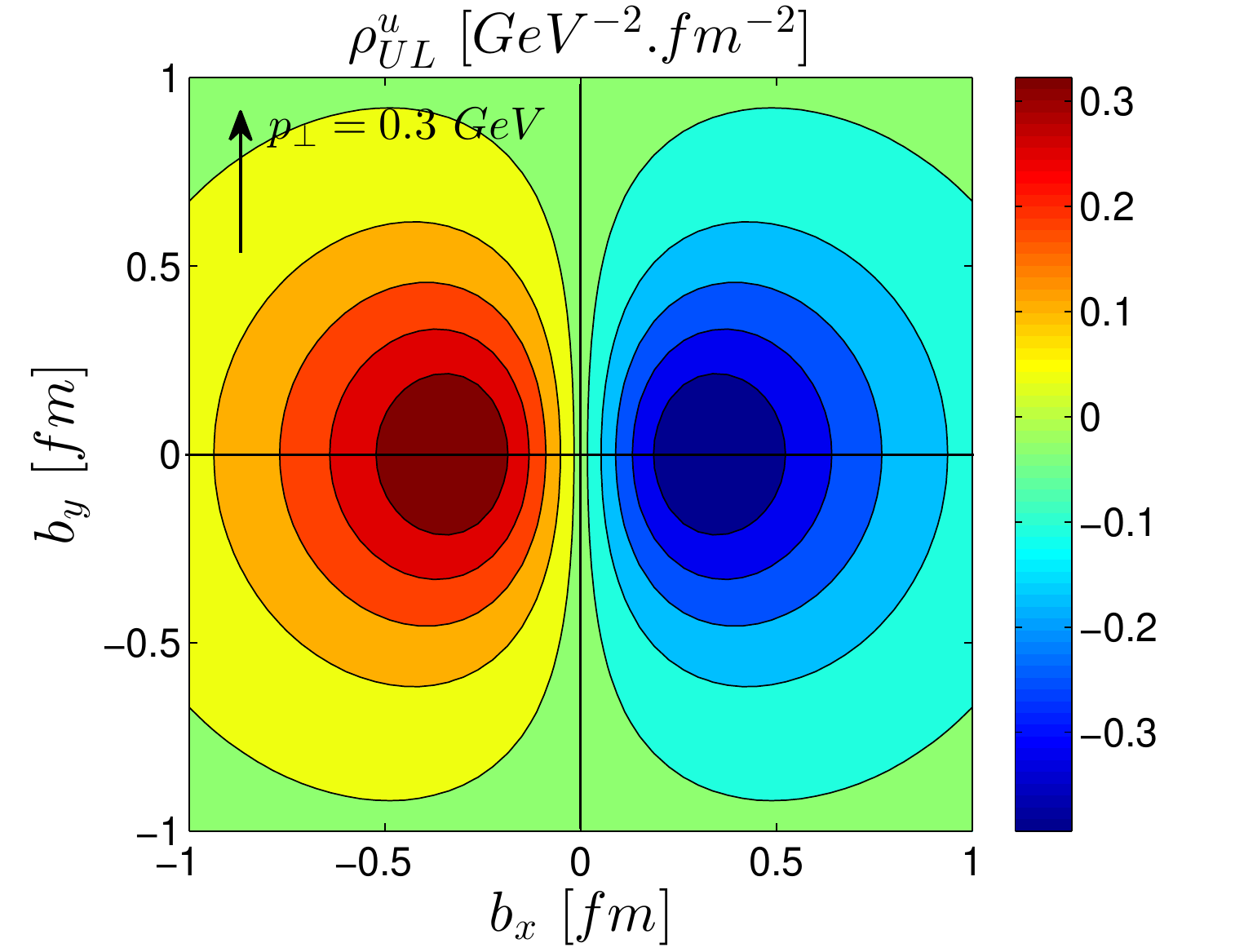}\\[5ex]
 \caption{Plots of $u$-quark Wigner distributions $\rho_{UU}$ and $\rho_{UL}$ in $b$ space
 in light-front quark-diquark model \cite{dipankar1}. }
\end{figure}

\section{Conclusion}

We presented some recent advances in the calculation  of the Wigner distributions 
of quarks in dressed quark target model and in a quark-diquark model for the
proton using a modified LFWF obtained using Ads/QCD. The dressed quark target 
represents a composite spin
$1/2$ state with a gluonic degree of freedom, and we have calculated the
gluon Wigner distributions as well. These are
expressed in terms of overlaps of LFWFs and using their analytic form we
have calculated them. Better convergence of the results is obtained using
Levin's method of integration. Further work would involve inclusion of the
gauge link in their description.

\section{Acknowledgement}

AM thanks the organizers of the QCD Evolution Workshop at Jefferson Lab for
the invitation and support. J. More, S. Nair, D. Chakrabarti, C. Mondal and
T. Maji are thanked for collaboration.

\end{document}